\documentclass[sigconf]{acmart}
\renewcommand\footnotetextcopyrightpermission[1]{}
\setcopyright{none}
\usepackage{hyperref}
\usepackage[hyphenbreaks]{breakurl}
\usepackage{balance}
\usepackage{url}
\usepackage{color}
\usepackage{caption}
\usepackage{diagbox}
\usepackage{subfigure}
\usepackage{multirow}
\usepackage{booktabs}
\usepackage{epsfig,endnotes}
\usepackage{enumitem}
\usepackage{multirow}
\usepackage{booktabs}
\newcommand{\RNum}[1]{\uppercase\expandafter{\romannumeral #1\relax}}

\newcommand{\para}[1]{{\vspace{2pt} \noindent \textbf{#1}
    \hspace{6pt}}}

\definecolor{applegreen}{rgb}{0.55, 0.71, 0.0}

\newcommand{\etal}{{\em et al.\ }}
\newcommand{\eg}{{\em e.g.,\ }}

\newcommand{\secspace}{\vspace{-0.05in}}

\newcommand{\ad}[1]{{$\mathcal{A}$}}
\newcommand{\service}[1]{{$\mathcal{S}$}}

\newenvironment{packed_itemize}{
\begin{list}{\labelitemi}{\leftmargin=0.5em}
  \setlength{\itemsep}{1pt}
  \setlength{\parskip}{0pt}
  \setlength{\parsep}{0pt}
  \setlength{\headsep}{0pt}
  \setlength{\topskip}{0pt}
  \setlength{\topmargin}{0pt}
  \setlength{\topsep}{0pt}
  \setlength{\partopsep}{0pt}
}{\end{list}}

\begin{document}
\title{A Response to Glaze Purification via IMPRESS}

\author{Shawn Shan$^\dag$, Stanley Wu$^\dag$, Haitao Zheng, Ben Y. Zhao\\
$^\dag$ denotes authors with equal contribution\\
  {\em Department of Computer Science, University of Chicago}\\
  {\em \{shawnshan, stanleywu, htzheng, ravenben\}@cs.uchicago.edu}}

\begin{abstract}

  Recent work proposed a new mechanism to remove protective perturbation
  added by Glaze in order to again enable mimicry of art styles from images
  protected by Glaze.  Despite promising results shown in the original paper,
  our own tests with the authors' code demonstrated several limitations of
  the proposed purification approach.  The main limitations are 1)
  purification has a limited effect when tested on artists that are \textit{not
    well-known historical artists} already embedded in original training
  data, 2) problems in evaluation metrics, and 3) 
  collateral damage on mimicry result for clean images.  We believe these
  limitations should be carefully considered in order to understand real
  world usability of the purification attack.

\end{abstract}

\maketitle

\secspace
\section{Introduction}
\label{sec:intro}

Cao~\etal~\cite{cao2023impress} recently proposed a purification-based method (IMPRESS) to 
remove Glaze protection~\cite{shan2023glaze} that are designed to protect art
styles from AI mimicry. We first summarize how Glaze and IMPRESS work in \S\ref{sec:back}. 
Then we show IMPRESS achieve limited effectiveness against
Glaze in a variety of real-world mimicry scenarios. In short:

\begin{packed_itemize}
\item IMPRESS has weaker performance on artists who are not historical
  artists already with a large presence in the pretrained models;

\item IMPRESS works poorly on smooth surface art styles (\eg realism);

\item IMPRESS damages image quality even for clean images. 

\end{packed_itemize}

\secspace
\section{Background}
\label{sec:back}

Here, we first summarize Glaze and then IMPRESS purification method. 

\para{Glaze protection against style mimicry. } Glaze seeks to protect artist's 
artwork from AI mimicry by adding small 
perturbations on these artwork to confuse diffusion models. 
Given an artwork $x$ and target style $T$ that is different from the artist's, 
Glaze first uses a pretrained style transfer model $\Omega$ to compute a style 
transferred version of the artwork. We denote such image as $\Omega(x, T)$. Then, Glaze 
computes a cloak $\delta_x$ that optimize the latent representation of Glazed artwork
($x + \delta_x$) to be similar to the style transferred artwork ($\Omega(x, T)$). 
The Glaze optimization effectively moves the original image to a new position in 
the high dimensional latent space, causing model to learn an incorrect art style. 
Glaze calculates the latent space using the 
feature extractor ($\mathcal{E}$) from a diffusion model.
Formally, we write the Glaze 
optimization as solving the following:
\begin{align}
    \min_{\delta_x} ||\mathcal{E}(x + \delta_x) - \mathcal{E}(\Omega(x, T))||_2 \\
    \text{subject to } \text{LPIPS}(x + \delta_x, x) < p_{G} \notag
\end{align}
\noindent We use LPIPS, a popular human-perceived visual distortion metric~\cite{zhang2018unreasonable}, to bound the perturbation 
within a budget $p_{G}$. 

\para{IMPRESS Purification Method. } IMPRESS adds additional perturbation on top of a Glazed artwork 
hoping to ``purify'' the Glaze
effect -- recovering the precise latent representation of original artwork. 
First, the authors empirically find that when passing Glazed images through an image autoencoder, 
the output image looks more different from the input image, compared to the output 
when inputting a clean image to the same autoencoder. Then authors assume removing this particular 
discrepancy would guide them to find the original (non-Glazed) image. 

IMPRESS purification  
optimizes perturbations on Glaze images such that purified images
behave similarly to clean images
when passing through an autoencoder. The authors assume the optimization process will guide 
the image to move back to the original latent space of the non-Glazed image. 
Formally, IMPRESS purification optimize a perturbation $\delta_{pur}$ on 
a Glaze image $x_{glazed}$:

\begin{align}
    \min_{\delta_{pur}} ||(x_{glazed} + \delta_{pur}) - \text{VAE}(x_{glazed} + \delta_{pur}) ||_2^2 \\
    \text{subject to } \text{LPIPS}(x_{glazed} + \delta_{pur}, x_{glazed}) < p_{I} \notag
\end{align}

\noindent $\text{VAE}$ is an image autoencoder, which consists of an encoder $\mathcal{E}$ followed by a decoder $D$. IMPRESS uses the same
autoencoder as the stable diffusion model. The 
perturbation $\delta_{pur}$ is bounded by a LPIPS 
perturbation budget $p_{I}$. 

\secspace
\section{Evaluation of IMPRESS}
\label{sec:issues}

Here, we first identify a few flaws in IMPRESS's experiment setup 
and then evaluate purification performance 
in several realistic mimicry scenarios. 

\para{Setup. } We follow the purification setup from the original IMPRESS paper and the source
code provided by the authors. We use default 
Glaze setup (perturbation budget = $0.07$) as described in the paper. 
We follow style mimicry setup in prior 
work~\cite{shan2023glaze}. Given a set of artworks, we mimic its style by
fine-tuning stable diffusion model (version 1.5) using DreamBooth 
for 600 to 1000 steps depending on the finetuning sample size for the artist.

\begin{figure*}
  \centering
  \includegraphics[width=0.99\linewidth]{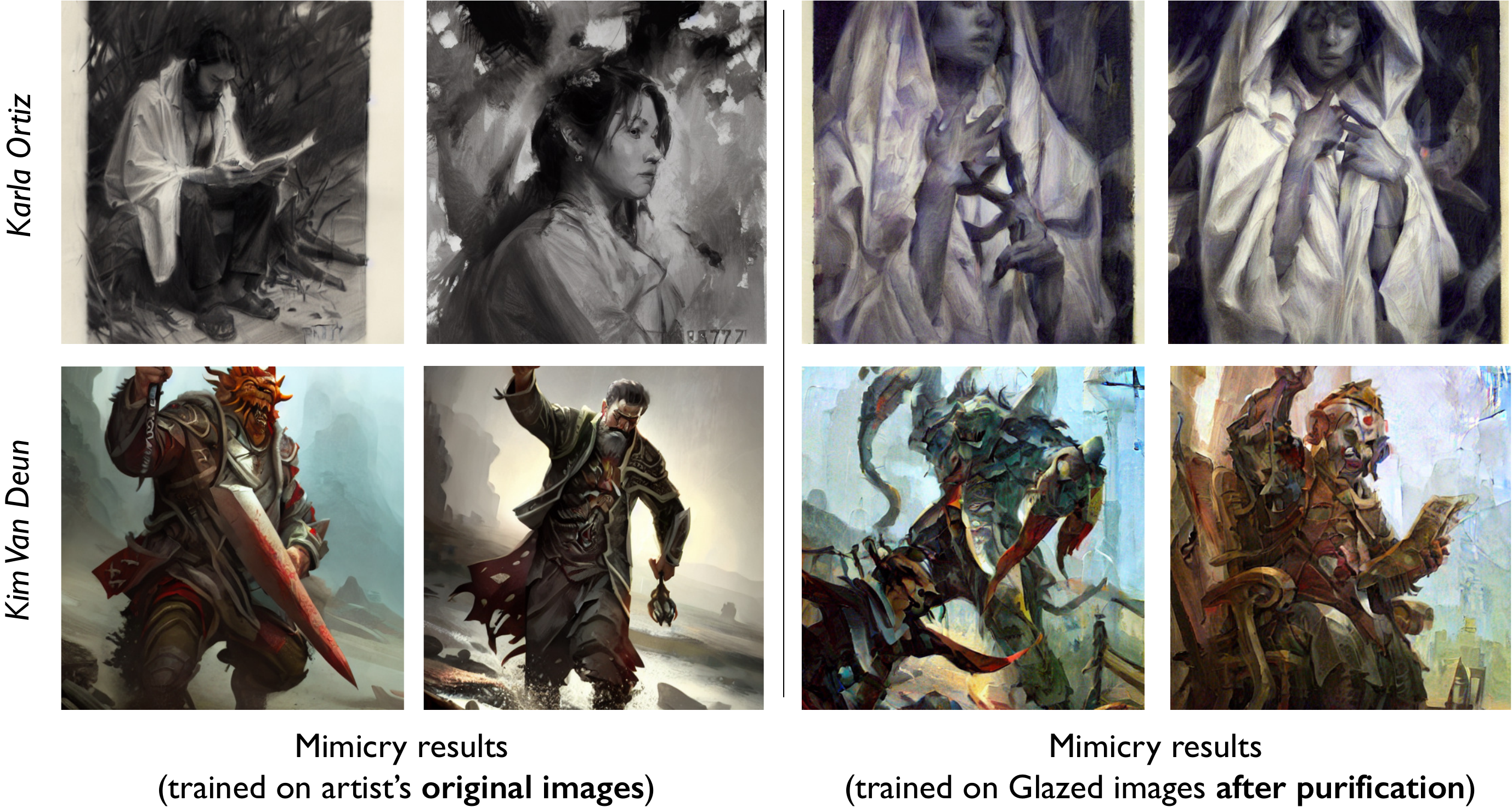}
  \vspace{-0.1in}
  \caption{Mimicry results on non-historical artists (Karla Ortiz and Kim van Deun). Left shows the images generated from a model trained on original images; right shows the images generated from a model trained on images that are first Glazed and then purified by IMPRESS. }
  \label{fig:non-historical}
  \vspace{-0.in}
\end{figure*}

\begin{figure*}
  \centering
  \includegraphics[width=0.99\linewidth]{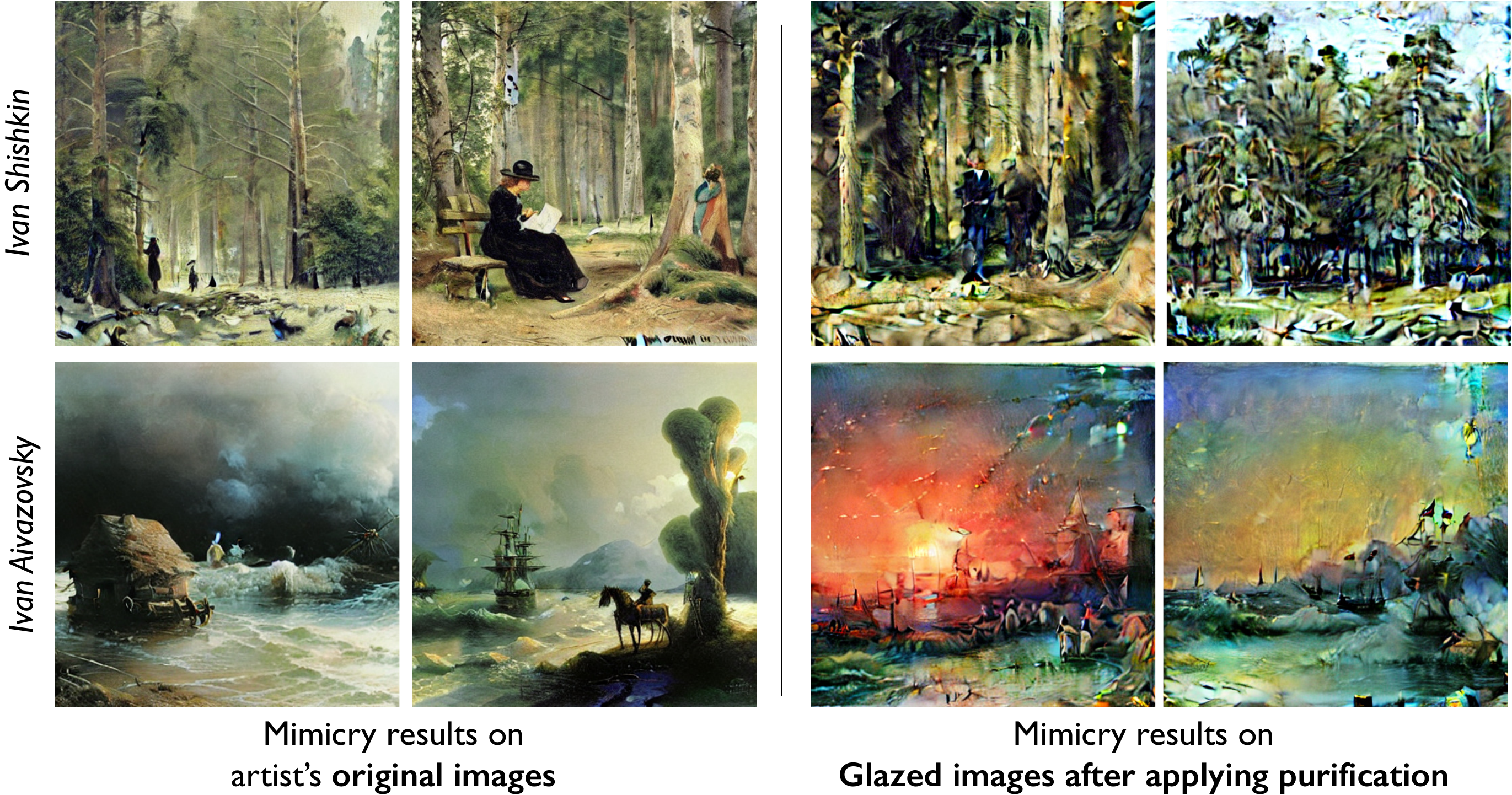}
  \vspace{-0.1in}
  \caption{Mimicry results on smooth surface art styles (Ivan Shishkin and Ivan Aivazovsky). Left shows the images generated from a model trained on original images; right shows the images generated from a model trained on images that are first Glazed and then purified by IMPRESS. }
   \label{fig:smooth}
  \vspace{-0.in}
\end{figure*}

\begin{figure*}
  \centering
  \includegraphics[width=0.95\linewidth]{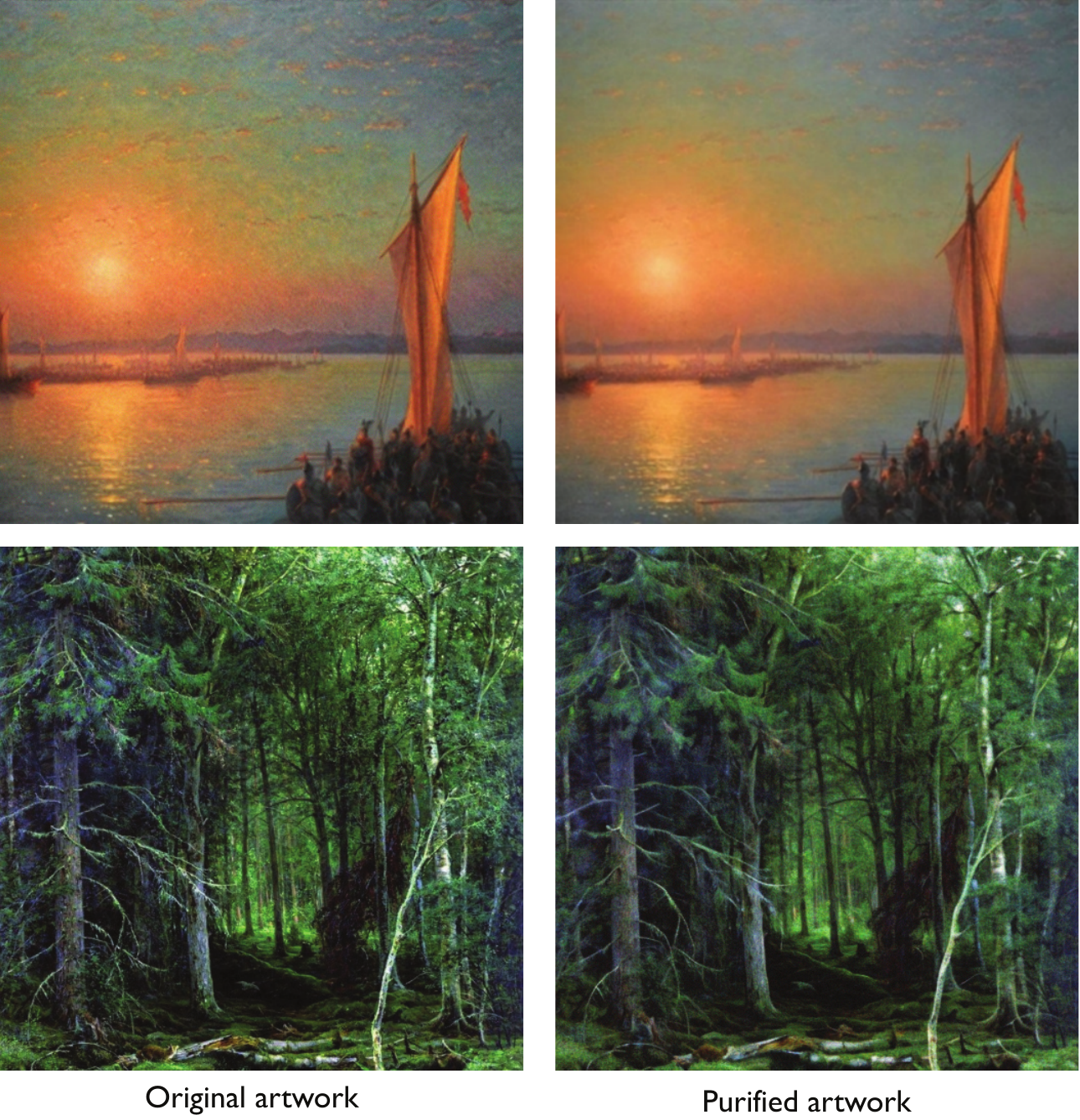}
  \vspace{-0.1in}
  \caption{Original artwork and corresponding purified artwork. }
  \label{fig:clean}
  \vspace{-0.in}
\end{figure*}

\subsection{Generalizability to Realistic Scenarios}

In the original IMPRESS paper, the authors focus the evaluation on protecting the
art styles of famous historical artists (Monet, Van Gogh) -- whose style are already learned
by pretrained diffusion models prior to style mimicry. In the real-world, it is current artists
who are most concerned about AI mimicking their art style. Glaze is designed
to protect those artists, and they are not as heavily pretrained into the
base model as Monet or Van Gogh. 

We evaluate IMPRESS on non-historical artists and show that purification
has limited effectiveness at removing Glaze protection. 
Even for historical artists, we find the purification works poorly for 
certain art styles. Lastly, we find purification also degrades clean image quality where
it removes texture from art pieces. 

\para{Performance on non-historical artists. } We use artwork from Karla 
Ortiz and Kim Van Deun to simulate the mimicry attack on current artists. Karla is a fine-art artist 
with a similar style as some historical artists tested in the IMPRESS paper. 
Figure~\ref{fig:non-historical}
shows the mimicry results when model trained on artist's original art pieces (left)
or when model trained on art pieces that are first Glazed
and then purified by IMPRESS (right). We can observe significant amount of 
artifacts on mimicry results when the model is trained on purified Glazed images. 

\para{Poor performance on certain art styles. } IMPRESS works by adding artifacts
on Glazed images to recover the latent representation of original artwork. 
We found purification has more challenges recovering smooth surface 
art styles (realism art, symbolism, romanticism, etc) even for historical
artists already trained into the base model. 
We choose two historical artists: Ivan Aivazovsky (romanticism style) and Ivan Shishkin (realism style). 

Figure~\ref{fig:smooth} shows the mimicry results. We see IMPRESS 
introduces signficiant amount of artifacts to the mimicked images. 
The weaker performance is likely because purifying  the 
original smoother surface art requires the optimization to 
be very percise -- find the exact smooth surface. 

\para{Degrading image quality. } We found the purification process degrades 
clean image quality. Figure~\ref{fig:clean} shows original artwork 
and its corresponding purified artwork. The purified artwork is much more
blurry as purification removes textures from the images. 

\subsection{CLIP-based metrics are Inaccurate}

``CLIP genre accuracy'' quantifies if mimicked art is classified into the same art genre as the original 
art pieces according to a CLIP model. It has been used in prior work to evaluate mimicry 
success. However, in our own tests dating back to late September 2023, we
found CLIP genre accuracy is a poor indicator of end-to-end mimicry success.  
CLIP accuracy is especially poor when evaluating attacks against Glaze. The 
reason is that attacks (such as 
IMPRESS purification), as they seek to remove Glaze effect from art, often have signficant impact 
on the base image quaility of the artwork. 
The degradation in image quality is not captured by CLIP accuracy, \eg a very
low quality cubism painting is still classified as ``cubism'' with high
probability. But the result are not successful 
mimicry due to the low quaility of the mimicked images. 
Because of these poor properties as an accuracy indicator, we stopped using
CLIP distance as a success metric starting with the Glaze v1.1 update
(October 2023). 

\secspace
\section{Discussion}
\label{sec:discuss}

We conclude with a discussion of Glaze, then describe some potential implications of 
attempts to remove Glaze. 

\para{Challenges of evaluating mimicry protection. } Evaluating mimicry performance is
challenging with automated metrics. As we shown in \S\ref{sec:issues}, CLIP-based metric
do not work well. Another popular image quaility metric, Fréchet inception 
distance (FID), is also shown to be erronous at evaluating generative models~\cite{podell2023sdxl}. 
The standard evaluation approach today is to rely on human inspectors~\cite{podell2023sdxl}, 
which is not only expensive, but also produces results with high
variance. Particularly for art mimicry, 
the average human user (found on MTurk or Profilic) often does not have sufficient expertise
to judge whether a mimicry is successful or not, while sourcing to professional artists
can be prohibitively expensive. We hope future work can address this challenge by
designing automated metrics that can faithfully approximate ratings from human
artists. 

\para{Implications on Copyright Law.} Technical research results and their applications rarely have
legal implications in practice. In the case of Glaze and image copyrights,
however, there are interesting open questions with respect to copyright
law. More specifically, Section 1201 of US Copyright law (enacted in 1998 as
part of the DMCA)  ``prohibits circumvention of technological protection
measures employed by or on behalf of copyright owners to control access to
their works.''\footnote{``Legal Liability for Removal or Alteration of Copyright Management Information Under the 
DMCA,'' John DiGiacomo, September 2020, 
\url{https://revisionlegal.com/copyright/legal-liability-for-removal-or-alteration-of-copyright-management-information-under-the-dmca/}.}

For artists who apply Glaze to prevent unauthorized use of their art
images for training, applying Glaze to one's art likely qualifies as an
explicit statement of ``opt-out'' and a measure to control access to
copyrighted work. In this uncertain legal and regulatory landscape, it is an
interesting legal question to explore whether Section 1201 applies to the use
of Glaze and attempted counter-measures to Glaze.

\section*{Acknowledgements}  
We are grateful to Bochuan Cao, Changjiang Li, Ting Wang, 
Jinyuan Jia, Bo Li, Jinghui Chen for providing us code and 
discussing their attack with us prior to paper publication.

\bibliographystyle{ACM-Reference-Format}
\bibliography{impress}
\balance

\end{document}